# CVD grown bilayer MoS$_2$ based artificial optoelectronic synapses for arithmetic computing and image recognition applications

Umakanta Patra[1], Subhrajit Sikdar[1,2], Roshan Padhan[3], Amandeep Kaur[1], Satyaprakash Sahoo[4,5], Subhabrata Dhar[1*]

[1]Department of Physics, Indian Institute of Technology, Bombay, Powai, India.

[2]International Center for Materials Nanoarchitectonics (MANA), National Institute for Materials Science, Tsukuba, Ibaraki 305-0044, Japan.

[3]Layered Materials and Device Physics Laboratory, Department of Chemistry, Physics and Atmospheric Science, Jackson State University, Jackson, MS 39217, USA.

[4]Laboratory for Low Dimensional Materials, Institute of Physics, Bhubaneswar, Odisha 751005, India.

[5]Homi Bhabha National Institute, Training School Complex, Anushakti Nagar, Mumbai, 400094, India.

*Corresponding author email: dhar@phy.iitb.ac.in

**Abstract**

Demand for lower computing power has rapidly increased. In this context, brain-inspired neuromorphic computing, which integrate data storage and processing, has attracted significant attention. Here, our study reveals that field effect transistors fabricated on chemical vapor deposited bilayer (2L) MoS$_2$ films can mimic the functions of biological synapse. These devices demonstrate high level of pair pulse facilitation (PPF), short term to long term memory (STM-to-LTM) transition as well as learning-forgetting-relearning properties. Effect of light intensity, pulse number, pulse width and photon energy on the STM-to-LTM transition is studied. It has been found that the rate of depression of the memory state can be controlled using the gate bias. Electrical and optical energy consumptions per synaptic event are estimated to be as low as 280 fJ and 20 nJ, respectively. Furthermore, photocurrent in these devices is observed to increase linearly with the number of the excitation pulses. This property has been exploited to demonstrate different arithmetic operations by the device. Moreover, these devices show great potential for image recognition. Artificial neural network simulation has returned an image recognition accuracy of ~85%. All these findings show a great prospect of 2L-MoS$_2$ for developing low power, transparent and flexible neuromorphic devices.

1. Introduction

The increasing demand for lower computational power and higher data transfer speed between memory and processing units impose the limitation in conventional von Neumann architecture, where the memory and computational units are physically separated.[1,2] There, the main bottleneck is the data bus, which limits the operational speed and power consumption efficiency while dealing with real time processing of complex tasks, such as speech, image and video that involve a large amount of data set.[3,4] Human brain, in contrast, is capable of performing much more complex operations with extremely low consumption of power[1-100 fJ per synaptic operation].[5] Researchers have identified that an artificially developed brain-like computer with integrated processing and memory units can potentially replace conventional von-Neumann architecture to meet the 'low power-high efficiency' requirement for the next-generation 'big data'-related computing.[6–8]

Artificial synapses with optical stimulation have been a superior choice in terms of achieving faster transmission speed, wider bandwidth and reduced crosstalk phenomena.[9,10] Two terminal memristive devices have potential for optoelectronic artificial synapses, where two electrical contacts resemble pre- and post-synaptic neurons.[11–13] However, in these devices, signal transmission and leaning-relearning process occur asynchronously due to the presence of only two terminals. In addition, it also suffers with instability, secondary data storage and inherent randomness.[14–16] In this regard, multi-terminal devices provide an advantage of simultaneous occurrence of signal transmission and self-learning processes.[17,18] Also, such devices offer higher stability, repeatability and better control over its transport properties.[19] Researchers have already started to mimic various functions of biological synapses by fabricating three terminal field effect transistor (FET) based artificial synaptic devices.[20–23]

Over the years, extensive research on two dimensional (2D) materials has demonstrated their ability to replace Silicon as the workhorse for the future electronic industry. Atomically thin structure of 2D materials enables high density of integration and flexibility.[24–29] Also, the layered structure leads to large surface to volume ratio which enhances their sensitivity to the external stimuli, which is extremely important for the operation of an artificial synapse. Synaptic devices based on 2D-transition metal dichalcogenides (TMDCs) have indeed been investigated by several groups.[30–32] It should be noted that the best performing devices are mostly reported on exfoliated samples. Molybdenum disulfide ($MoS_2$) is one of the most widely researched TMDC materials. Growth of continuous mono- and a few-layer $MoS_2$ films covering large area has also been achieved.[33,34] There are indeed reports of synaptic devices fabricated on chemical vapor deposition (CVD) grown mono- or few-layers of $MoS_2$. Shen et al. have reported two terminal synaptic devices on CVD grown monolayer $MoS_2$ island with a lateral size of ~ 50 $\mu$m [35]. The device shows an electrical power consumption of 1.8 pJ per synaptic event. Zhang et al. have demonstrated artificial synaptic array on CVD grown large area covered 1L-$MoS_2$ film[36]. However, the power consumption (electrical or optical) values are not provided. Synaptic devices based on CVD grown trilayer $MoS_2$ islands have been reported by Nam et al., which shows an electrical power consumption of ~6 pJ per synaptic event [37]. However, more are needed to be done to understand the prospect of large area covered $MoS_2$ films for the artificial synaptic device fabrication.

Note that synaptic response of a material relies on the persistent photoconductivity (PPC) effect[38]. PPC depends on specific type of trap formation in the system[39]. These traps are often associated with defects/adsorbates present at the surfaces/interfaces of a material[40]. Unlike monolayer $MoS_2$, bilayer (2L) and multilayer-$MoS_2$ offer interfaces between the layers, which can be intercalated with adsorbates. This can open up the chance of formation of trap states that are unique only to the intercalation. Such traps may lead to the PPC effect different from that is found in 1L-$MoS_2$. It is also plausible that a gate bias in a transistor structure can act on the intercalated adsorbate species to modify the potential landscape of the traps. This can create a unique scenario of electric field-controlled filling of the traps resulting in the tunability of the PPC characteristics[41]. Moreover, the PPC characteristics of a bilayer is supposed to be dominated by the intercalated traps. The synaptic performance of a bilayer device can thus be less perturbed by the surroundings as compared to the monolayer-based devices. Note that higher density of traps may result in much longer PPC effect in multilayer-$MoS_2$, which will lead to the difficulty in resetting the synaptic weights[42]. Considering a three-terminal device, bi-layer structure offers better electrostatic gate controllability on the occupation of the traps. In multilayer devices, it is difficult to access traps residing deep inside. This accessibility is directly related to the programmability of synaptic weights. A bilayer device should thus be a good balance between PPC timescale, strength and its controllability.

Here, we experimentally demonstrate that bilayer MoS$_2$ based FETs can act as optoelectronic synapses with pair pulse facilitation (PPF) index of as high as 135%. These devices are also shown to replicate the short term to long term memory (STM-to-LTM) transition as well as learning-forgetting-relearning behaviour of the human brain. Synaptic performance of these FETs is systematically studied as functions of different parameters associated with the external optical stimuli, such as pulse number, pulse width, light intensity and photon energy. It has been found that the rate of depression of the memory state can be controlled using the gate bias. These devices are estimated to consume as low as 280 fJ and 20 nJ of electrical and optical energy, respectively per synaptic event. These numbers are comparable with the best of the values reported for such devices in the literatures. These artificial synapses are also shown to perform arithmetic and image recognition operations. The device also shows a great prospect for image recognition. Artificial neural network (ANN) simulation of these devices shows an image recognition accuracy of as high as ~85%.

**Experimental section**

**Growth of bilayer MoS$_2$**

Samples were grown in a CVD chamber as depicted schematically in supporting information Figure S1. A ceramic boat containing 2 mg of Molybdenum Trioxide (MoO$_3$) powder (99.995 %, Sigma Aldrich) and another ceramic boat with 350 mg of Sulphur (S) powder (99.95%, Sigma Aldrich) were kept, respectively, in the hotter and cooler zones of the quartz reactor tube. A SiO$_2$/Si substrate was cleaned in Trichloroethylene, Acetone and Methanol for ten minutes each in an ultrasonic bath. The furnace temperature was raised to 150°C and Ar-purging was pursed for 20 minutes at $\varphi_{Ar} = 200$ sccm. $\varphi_{Ar}$ was then reduced to 8 sccm and the furnace temperature was raised to 700°C at a rate of 10°C/min. This temperature was maintained for 10 mins. At this condition, the S containing boat was kept at a temperature of ~ 150°C. Furnace was then allowed to cool naturally to RT. Note that details about the cavity arrangement for the growth of 2L-MoS$_2$ by using CVD technique has been discussed in our previous report.[43] After the growth, the layers are transferred to SiO$_2$ coated Si substrates using a polystyrene based transferred method.[44]

**Characterization of the Bilayer MoS$_2$ thin films**

Optical and atomic force microscopy (AFM) images were recorded at room temperature. Photoluminescence (PL) and Raman spectroscopy were carried out in a micro-PL/Raman setup [from Renishaw, Invia Reflex, UK] equipped with a 50× magnification objective lens. The sample was excited with 532 nm laser at 500 μW power. A 500 cm focal-length monochromator equipped with 1800 gr/mm grating and CCD detector was used to record the spectra. High resolution transmission electron microscopy (HRTEM) images were recorded using a 300 kV system from Thermo Scientific, Themis 300 G3 after transferring the film on a Cu-grid by polystyrene based wet transfer technique.

**Device fabrication and Characterization**

2L-MoS$_2$ based backgated FETs were fabricated by using standard optical lithography techniques. Deposition of Ti (30 nm) and Au (100 nm) was performed using sputtering technique. Current-Voltage (I-V) characteristics was measured by a 6487 Pico ammeter-voltage source from Keithley and 2B2900 source measuring unit (SMU) from Keysight. A set up was developed in the lab to illuminate the devices with lights of different wavelengths and powers using a Xenon lamp, monochromator and other necessary apparatuses.

2. **Results and Discussion**

Figure 1(a) shows the optical microscopy image of large area covered CVD grown MoS$_2$ film after transferring on a SiO$_2$ coated Si substrate by polystyrene based wet transfer technique.[44]

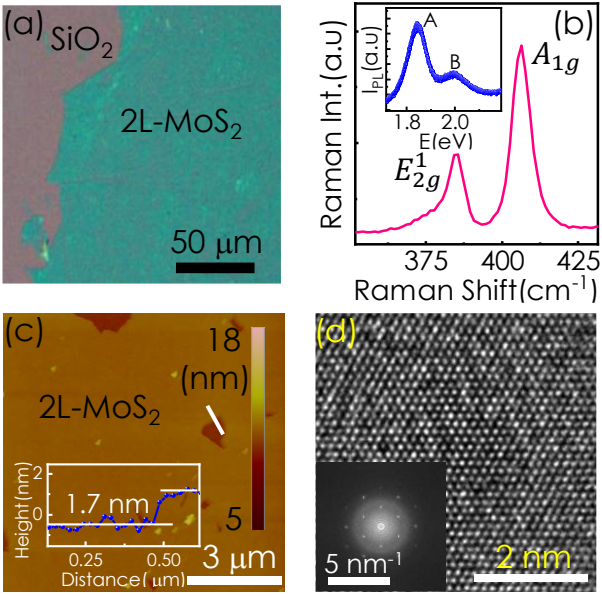
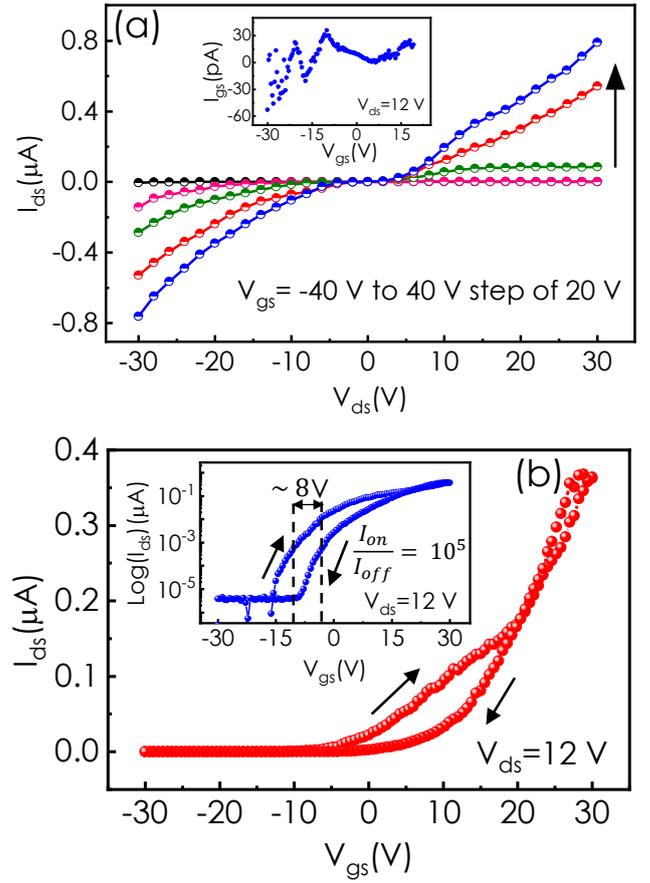

**Figure 1.** (a) Optical microscopy image showing a large area covered 2L-MoS$_2$ film after transferring on to a SiO$_2$/Si substrate. (b) Plot of room temperature Raman spectra and the inset shows PL spectra taken on a 2L-MoS$_2$ film. (c) AFM image showing the continuous 2L-MoS$_2$ film and the inset shows height profile taken along the white line marked on the image. (d) HRTEM image of the 2L-MoS$_2$ film with SAED pattern at the inset confirms the growth of highly crystalline 2L-MoS$_2$ film.

Room temperature Raman spectrum is presented in Figure 1(b). The spectrum is featured by two peaks, which can be identified as the $E^1_{2g}$ and $A_{1g}$ phonon modes associated with the in-plane and out-of-plane vibrations in MoS$_2$. The two peaks are estimated to be separated by 21 cm$^{-1}$, which matches very well with the reported values for the bilayer MoS$_2$ films.[45] Inset of Figure 1(b) shows the room temperature PL spectrum of the layer. The peaks located at 1.84 eV (marked as A) and 2 eV (marked as B) can be attributed to the neutral A- and B-excitons, respectively. Figure 1(c) presents the AFM image of the MoS$_2$ layer. Thickness of the film has been estimated as 1.7 nm from the line scans recorded at the edges of the film (as shown in the inset). The value matches very well with that is expected for bilayer MoS$_2$ films.[46] Figure 1(d) shows the HRTEM image of one of the grown films and the corresponding selective area electron diffraction (SAED) pattern (in the inset). The diffraction pattern confirms the growth of highly crystalline MoS$_2$ films.

**Figure 2.** (a) Output characteristics of the FET device for different $V_{gs}$. Inset of (a) gate leakage current profile (b) The transfer profile of the device recorded at $V_{ds} = 12$ V. Inset of (b): the transfer profile in log-scale.

Figure 2(a) shows the output characteristics for different $V_{gs}$ values. Gate-source current $I_{gs}$ (Gate leakage current) is plotted as a function of $V_{gs}$ in the inset of figure 2(a). Gate leakage is found to be of the order of only a few tens of pA. Dual sweep $I_{ds} - V_{gs}$ profile is shown in figure 2(b). Our 2L-MoS$_2$ based back-gated FET device shows a hysteresis window of ~8V. Threshold voltage is obtained by plotting $\sqrt{I_{ds}}$ as a function of $V_{gs}$ [see the supporting information figure S2]. $V_{th}$ is found to be $-10.2$ V from the intercept (crossing point with $y = 0$ line). Electron concentration ($n$) has been calculated by using the expression [47]

$$n = C_{ox} \times (V_{gs} - V_{th})/e \quad (1)$$

Where $e$ is the electronic charge, $C_{ox}$ is the gate oxide capacitance per unit area. It has been estimated from the expression

$$C_{ox} = \varepsilon_o \varepsilon_r / d \quad (2)$$

Where $\varepsilon_o = 8.85 \times 10^{-12}$ is the permittivity of the free space, $\varepsilon_r = 3.9$ the dielectric constant of SiO$_2$ and $d = 300$ nm is the thickness of the SiO$_2$ layer. $C_{ox}$ is estimated to be 11.5 nF/cm$^2$. Background electron concentration (at $V_{gs} = 0$ V) is found to be $7.3 \times 10^{11}$ cm$^{-2}$. The carrier mobility has been estimated from the transconductance $g_m =$

$$dI_{ds}/dV_{gs} \quad (3)$$

following the relation[48]

$$\mu = g_m L / C_{ox} W V_{ds} \quad (4)$$

Where $L$ (~ 10 μm) is the channel length, $W$ (~ 100 μm) is the channel width. Electron mobility $\mu$ of the layer has been obtained as 0.03 cm$^2$V$^{-1}$s$^{-1}$. On/off ratio of the device is found to be ~$10^5$ [see the inset of Figure 2(b)].

Figure 3(a) shows the Ti/Au source and drain contacts of the fabricated FET where red lines define the area covered by 2L-MoS$_2$. The resemblance between the device and a biological synapse is schematically depicted in Figure 3(b). The source-drain contacts play the role of the pre- and post-synaptic terminals of a synapse responsible for visual sensory, where the optical pulses are the external stimuli. Excitatory post synaptic current (EPSC) is measured between source and drain terminals ($I_{ds}$), which is modulated under optical pulses of different wavelengths. Figure 3(c) shows the $I_{ds}$ versus $V_{ds}$ loop recorded at $V_{gs} = 0$ V, which suggests a resistive switching (RS) behaviour. In the supplementary figure S3, the profile is shown in the log-log plot, where slope-changes at various locations can clearly be observed. Such changes of slope can be attributed to changeovers between different conduction mechanisms.[49,50] RS can be attributed to the migration of the sulphur vacancies playing the role of neurotransmitters in a real biological system. It has been speculated that the transition from high resistive state (HRS) to low resistive state (LRS) is resulting from the Sulfur defect mediated filament formation (SET) and rupture (RESET) processes.[51–53] In our devices, 2L-MoS$_2$ can provide an additional pathway for switching via free molecules/radicals intercalated between the two monolayers. These species can also be driven by the applied electric field to facilitate the formation of additional conductive filaments, which leads to the resistive switching behaviour. Figure 3(d) presents the photo response ($I_{ds}$, the EPSC) of the FET recorded at $V_{gs} = 0$ V and $V_{ds} = 10$ V after illuminating it with an optical pulse of width $\Delta w = 10$ s, wavelength 532 nm and power 20 μW. An abrupt increase of the photocurrent from 0.4 nA to 4.1 nA is observed under illumination due to an efficient generation-separation-collection process of photogenerated electron-hole pairs. When the light is switched off the photocurrent shows a gradual reduction. This persistent photoconductivity (PPC) effect can be attributed to the traps present on the surface and/or layer/dielectric interface[54]. The decay profile is fitted with two exponential curves as given by [55]

$$I_{ds} = I_o + C_1 e^{(\Delta t/\tau_1)} + C_2 e^{(\Delta t/\tau_2)} \quad (5)$$

where $I_0$ is the dark current, $C_1$, $C_2$ are the constants and $\tau_1, \tau_2$ are the decay time constants. The fitting returns $\tau_1$ and $\tau_2$ as 2.5 secs and 24 secs, respectively, which is typically reported in TMDC based synaptic devices[38,56]. It should be noted that the time constant for biological synapse is about 1-2 ms [57]. Note that we have performed EPSC response on multiple devices fabricated on the same 2L-MoS$_2$ film, which show very similar response characteristics. The results on two such samples are shown in supporting information figure S4.

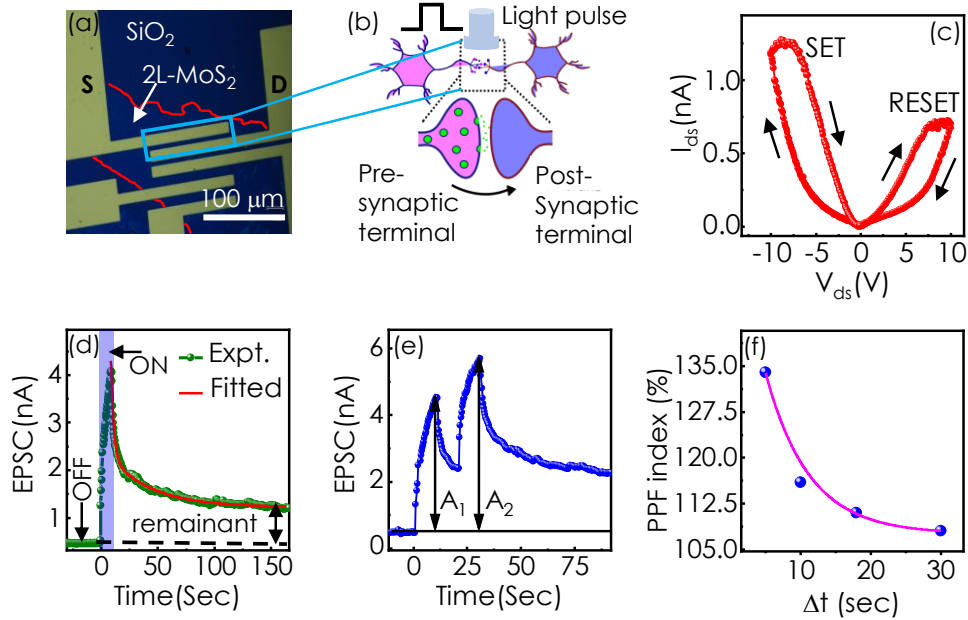

**Figure 3.** (a) Top view optical microscopy image of our fabricated back gated field effect transistor based on 2L-MoS$_2$ films, (b) Schematic representation of a FET acting as an artificial optoelectronic synapse, (c) $I_{ds} - V_{ds}$ characteristics under dark condition at $V_{gs} = 0$ V, (d) Photocurrent profile obtained at $V_{ds} = 10$ V and $V_{gs} = 0$ V under illumination of an optical pulse of width 10 secs and power of 20 μW, (e) PPF behaviour under the illumination of two consecutive optical pulses, (f) Variation of PPF index with time interval between two consecutive optical pulses.

Paired pulse facilitation (PPF) is the characteristics of short-term plasticity of a biological synapse which represents the collective effect of two consecutive presynaptic stimulation on the excitatory post synaptic current. PPF is crucial for decoding and recognition of the auditory and visual information.[58] PPF index is defined as the ratio of peak EPSC values attained due to second pulse ($A_2$) to first pulse ($A_1$),

$$\text{PPF index} = \frac{A_2}{A_1} \times 100\% \quad (6)$$

PPF effect is studied under the illumination of two consecutive optical pulses of $\Delta w = 10$ s and interval $\Delta t = 10$ s, when $V_{gs} = 0$ V and $V_{ds} = 10$ V as shown in Figure 3(e). PPF index is estimated to be as high as 115% at these conditions. The variation of the PPF index as a function of the time interval between the consecutive pulses $\Delta t$ is plotted in Figure 3(f). Evidently the index value gradually decreases from 134% to 105% as $\Delta t$ is increased from 5 to 30 s. The dependence of PPF on $\Delta t$ is in fact found to show an exponential behaviour. One can see from Figure 3 that the conductance of our back gated 2L-MoS$_2$ based artificial synaptic device can be modulated by the electrical and optical stimuli. The device can thus mimic the behaviour of a biological synapse.

Modulation of synaptic weight by presynaptic stimulus is essential for data accumulation and transformation in the neural network.[59] In the human brain, complex learning and information processing involves interaction among the multiple spatially and temporally correlated input stimulus as governed by Hebb's learning rule.[60] As per the learning-relearning-memorizing rule, short term memory (STM) which represents weak potentiation on the application of external stimulus and can be retained for a few seconds to a few minutes. However, repetition or deep learning of STM can bring a permanent change in the synaptic weight called long term memory (LTM), that can store information for hours, months and sometimes for the life time.[61] In our work, such STM to LTM transition is demonstrated in terms of varying the optical pulse number, pulse width, power and energy of the incident light. For these studies, FET is operated at $V_{gs} = 0$ V and $V_{ds} = 10$ V. Figure 4(a) represents STM to LTM transition with the number of optical pulses of width $\Delta w = 10$ s and interval $\Delta t = 10$ s. EPSC increases gradually from 0.4 nA to 7.8 nA as pulse number increases from 1 to 3. Figure 4(b) shows the EPSC response profiles for different values of the pulse with $\Delta w$. One can see more than an order of magnitude increase of EPSC as the pulse width changes from 10 to 30 s. Such increase of EPSC leading to the transition from STM to LTM is due to two concurrent phenomena; (a) efficient generation and separation of photoexcited electron-hole pairs and, (b) the PPC effect. STM to LTM is also possible by increasing the illumination power as shown in Figure 4(c).

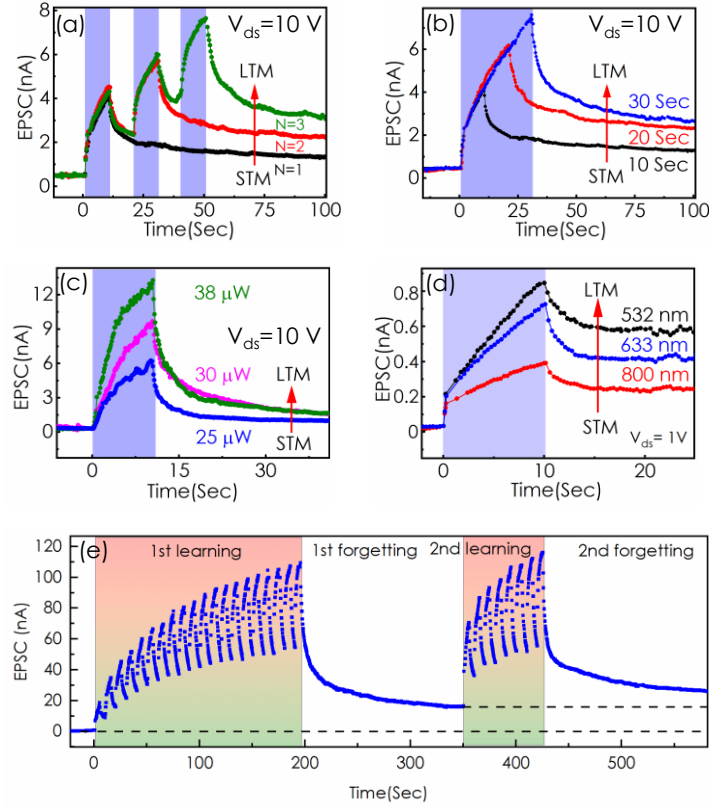

**Figure 4.** STM to LTM transition in our 2L-MoS$_2$ FET based artificial optoelectronic synapse with (a) pulse number, (b) pulse width, (c) illumination power, and (d) light wavelength. (e) Learning-forgetting-relearning behaviour of our 2L-MoS$_2$ based artificial optoelectronic synapse.

The device is also illuminated with lights of different photon energies at a photon power of 20 $\mu$W for $\Delta w = 10$ s. The time response profiles of EPSC are compared in Figure 4d for $V_{gs} = 0$ V and $V_{ds} = 1$ V. Evidently, the signal increases with the photon energy demonstrating an additional way to move from STM to LTM states. It should be noted that denoising is an important requirement in image processing. The property of STM to LTM transition can be utilize to perform denoising at the sensor itself once it is trained to memorize an image.[62] This process, which is termed as 'self-denoising', requires LTM to STM ratio to be more than 2 for efficient execution. As shown in Figures 4(a-d), this condition can be achieved in our synaptic device by varying any of the optical pulse parameters, the number, width, power and energy. Moreover, one can utilize the dependence of EPSC on the photon energy to denoise images from colour interference, mimicking the operation of the biological retina.[62] Self-denoising of image has indeed been demonstrated in an array of artificial optoelectronic synapses based on 2L-MoS$_2$.[62]

Human learning experience can be a combination of intermittent processes of learning, forgetting, relearning etc. We wanted to check how our artificial synapse can mimic the human learning experience. First, the device (under $V_{gs} = 0$ V and $V_{ds} = 10$ V) is illuminated with 20 pulses of 532 nm laser light ($\Delta w$ = 5 s, $\Delta t$ =5 s power = 20 μW).

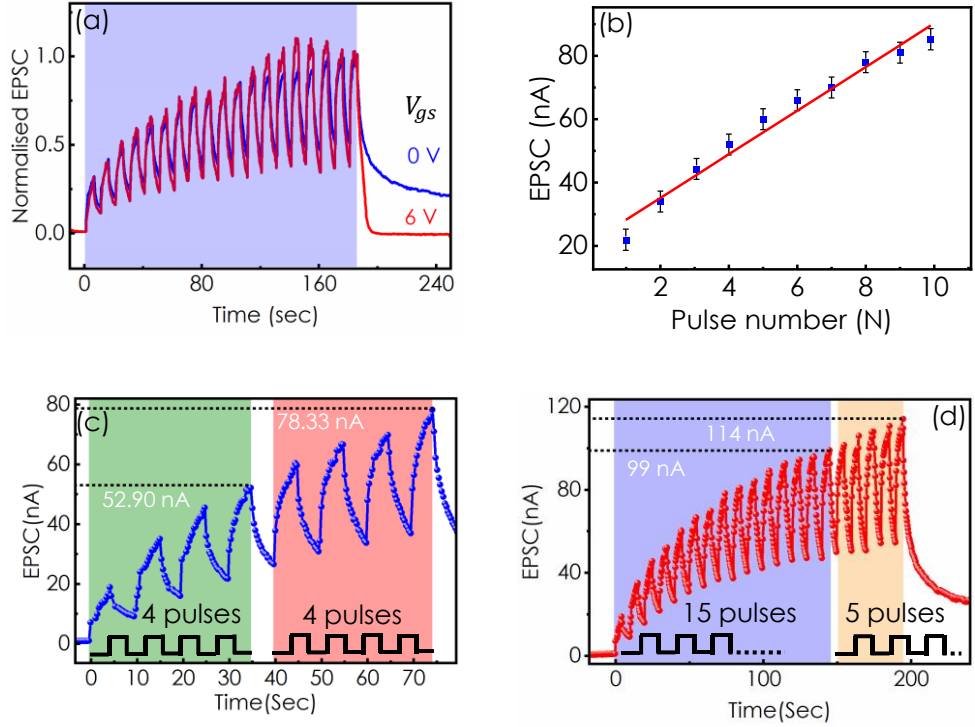

**Figure 5.** (a) Plots the depression of EPSC under 0 and +6 V gate biases. (b) EPSC signal as a function of pulse number $N$. Demonstration of (c) addition and (d) subtraction operations.

As shown in Figure 4(e), EPSC increases from 0.4 to 110 nA at the end of the 20$^{th}$ pulses. This state of EPSC can be treated as the state of the memory after learning. The light is then switched off and the EPSC is allowed to decrease to 16 nA, which can represent the state of the memory after the first forgetting. The device is then illuminated again. EPSC now takes only 8 pulses to reach that 110 nA value. This can represent the relearning process. The illumination is switched off again. Note that the EPSC reaches a higher value as compared to the 1$^{st}$ forgetting experience after the same amount of decay time. This means that the memory stays at a better state after the second forgetting process than the first one, which is due to the relearning experience the device goes through in-between. Note that the human memory also works in a similar fashion. Decay profile of EPSC for 1$^{st}$ and 2$^{nd}$ forgetting processes can be fitted with two exponential functions [see supporting information figure S5]. The two decay constants are found to be 3 s and 42 s for the 1$^{st}$ forgetting, while 2 s and 56 s for the 2$^{nd}$ forgetting processes. It is noteworthy that forgetting process of the human brain follow the Ebbinghaus forgetting curve, which also exhibits a faster initial decay followed by a slower rate of reduction.[32]

In an artificial synapse, it is worthy to have good control over both the potentiation and depression processes. Two terminal based optoelectronic devices face challenges to implement the controlled depression process and thus generally rely on spontaneous decrease of EPSC with time.[63] In this regard, three terminal devices provide an extra degree of freedom in the form of gate to modulate the EPSC whenever necessary.[37] Figure 5(a) shows the impact of positive gate voltage on the depression process of our device, which is kept under intermittent illumination ($\Delta w$ = 5 s, $\Delta t$ = 5 s) with 532 nm laser of power 20 μW for some time before completely switching off the light. During the illumination stage, $V_{ds}$ = 10 V and $V_{gs}$ = 0 V are maintained. While at the point of final switch-off of the light, the gate bias is either kept at $V_{gs}$ = 0 V (blue curve) or changed to $V_{gs}$ = 6 V (red curve). Evidently, the application of 6 V of positive gate bias at the light switch-off point results in a much faster reduction of the EPSC to the base value as compared to the zero-gate-bias case. It should be noteworthy that the EPSC has also been found to fall sharply to a much lower value whenever $V_{gs}$ = 6 V is introduced even under illumination (see the supporting information figure S6).

Controlled variation of synaptic weight plays a vital role for decimal arithmatic computing applications.[64] Linearity in EPSC with the change of pulse number forms the basis of such arithmatic computation. In order to explore this capibility of our synaptic device, we illuminate it with 20 light pulses of wavelength 532 nm, power 60 μW, $\Delta w$ = 5 s and $\Delta t$ = 5 s at $V_{ds}$ = 10 V and $V_{gs}$ = 0 V. Figure 5(b) shows the variation of ΔEPSC, the change of EPSC from its dark current value, with the pulse number $N$. Note that the cycle-to-cycle variation in EPSC is found to be ~ 3% [ see the supplementary materials figure S7]. The data can be fitted a straight line.

$$\Delta\text{EPSC} = 6.88\,N + 21.4 \qquad (7)$$

The addition operation is demonstrated in Figure 5(c). The device is first illuminated with 4 pulses which results in a ΔEPSC of 52.90 nA. Afterward, the device is exposed to 4 more pulses, which makes ΔEPSC to 78.33 nA. When the final ΔEPSC value is put into the equation 2, N comes out to be 8. The subtraction operation is demonstrated in Figure 5(d). First, the device is subjected to 20 optical pulses, which results in ΔEPSC = 114 nA. We take this value as the reference for the number 20. A depression voltage of 6 V is then applied to bring the EPSC to its base value. The device is again exposed to 15 pulses that leads to ΔEPSC = 99 nA. Now, it can be observed that only 5 (20 - 15 = 5) pulses are required to reach the reference current. Figure 5 thus demonstrates the decimal arithmetic operation in these devices. This opens up the possibility of using these artificial synapses for the development of a new type of computation, where the conventional binary logic can be mixed with the decimal arithmetic operations. Such a system can make calculations faster and more energy efficient.

Note that when our device is illuminated with 532 nm light pulse width $\Delta t$ = 400 ms and the power of 50 μW, the EPSC attains a peak value of 70 pA for $V_{gs}$ = 0 V and $V_{ds}$ = 10 mV as shown in the supporting information figure S8. This results in an electrical energy consumption ($E_{elec} = V_{ds} \times I \times \Delta t$) of 280 fJ per synaptic event and an optical energy consumption ($E_{optical} = P \times S \times \Delta t$) of 20 nJ per synaptic event. Note that the optical power density ($P$) and active area ($S$) of our device are estimated to be 0.01 μW/cm$^2$ and ~500 μm$^2$, respectively. Energy consumptions of the device per synaptic event is compared with the values reported for other MoS$_2$ based artificial synaptic devices in Table 1. It is noticeable that our device has the lowest electrical energy consumption per event among all the single and bilayer MoS$_2$ based devices reported in the literatures so far. Interestingly, Huh *et al.* reported a remarkably low electrical energy consumption of 0.1 fJ for a device fabricated on an exfoliated MoS$_2$ chunk of thickness more than 15 layers.[65] Note that in the device of Huh *et al.*, the MoS$_2$ chunk was placed on hexagonal boron nitride (h-BN) coated Si/SiO$_2$ substrate. h-BN is often used to eliminate the effects of interface traps between the MoS$_2$ channel and SiO$_2$/Si substrate. However, there are only a few reports of the optical energy consumption of the devices fabricated on MoS$_2$ platform. Huang *et al.* have reported the total (electrical plus optical) energy consumption of only 66.5 pJ for the device fabricated on CVD grown MoS$_2$ monolayer.[72] Note that, these devices are fabricated on the array of MoS$_2$ monolayers, which are grown with the gold nanorods as seeds. These metallic nanostructures can enhance the light coupling efficiency of these devices through exciton-plasmon coupling.

Table 1. Comparison of electrical and optical energy consumption values and image recognition accurecy of MoS$_2$ based artificial elelctronic and optoelectronic synapses.

| Material name | Growth technique | Layer number | Substrate | Electrical energy consumption | Optical energy consumption | Image recognition accuracy (%) | Ref. |
|---|---|---|---|---|---|---|---|
| MoS$_2$ | CVD | 1 | Si/SiO$_2$ | 1.8 pJ | --- | --- | [35] |
| MoS$_2$/MoS$_2$−xO$_\delta$ | CVD | 1 | Si/SiO$_2$ | 20 pJ | --- | --- | [66] |
| MoS$_2$ | CVD | 1 | Si/SiO$_2$ | 200 pJ | --- | 89.5 | [67] |
| MoS$_2$ | Sputtering | Multi-layer | Si/SiO$_2$ | --- | --- | 41 | [68] |
| MoS$_2$/Nb$_2$O$_5$ | CVD | 3 | Si/Al$_2$O$_3$ | 6 pJ | --- | 94.2 | [37] |
| MoS$_2$/ZnO | Exfoliation | 10 | Si/SiO$_2$ | 2.55 nJ | --- | --- | [69] |
| MoO$_3$/MoS$_2$/h-BN | Exfoliation | > 15 | Si/SiO$_2$ | 0.1 fJ | --- | 86 | [65] |
| MoS$_2$/MoO$_3$/WSe$_2$ | Exfoliation | 4 | Si/SiO$_2$ | --- | 6.9 pJ | --- | [70] |
| MoS$_2$ | Exfoliation | 40 | Si/HZO | --- | --- | 90 | [71] |
| MoS$_2$ | CVD | 1 | Si/HfO$_2$ | --- | --- | 83.1 | [36] |
| MoS$_2$ | CVD | 1 | Si/SiO$_2$ | --- | --- | 95 | [21] |
| MoS$_2$ | CVD | 2 | Si/SiO$_2$ | 280 fJ | 20 nJ | 85 | This work |

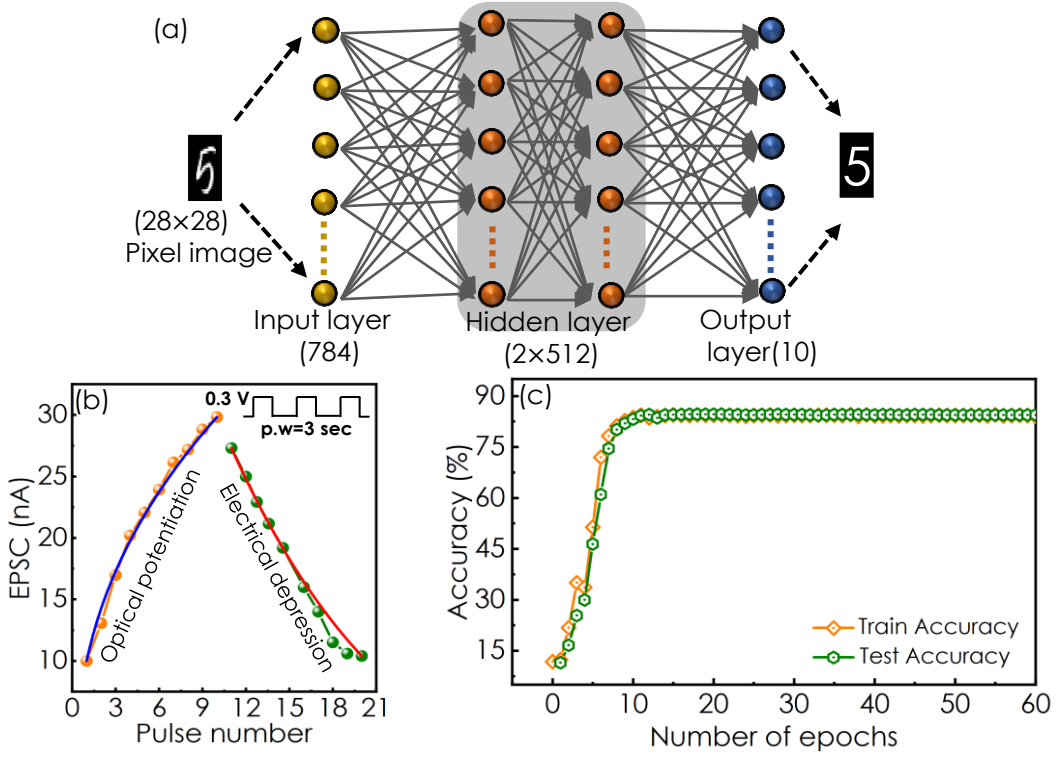

**Figure 6.** Handwritten image recognition performance of 2L-MoS$_2$ based synaptic device using an ANN, (a) Schematic of an ANN structure consisting of input (784 nodes), hidden (2×512 nodes) and output (10 nodes) layers used for image recognition, (b) Optical potentiation (10 optical pulses) and electrical depression (10 electrical pulses) of our devices with the corresponding non-linearity factors, (c) Plot of image recognition accuracy achieved with respect to the number of epochs from the ANN simulation.

To explore the potential of our artificial synapse for image recognition, we performed artificial neural network (ANN) simulations using the open-source software PyTorch. A three-layer ANN is constructed to train and test handwritten digits from the Modified National Institute of Standards and Technology (MNIST) database, as depicted in Figure 6(a). The network consists of input, hidden, and output layers with a structure of 784 × (2×512) × 10 nodes, where each 28×28-pixel handwritten images of different digits from "0" to "9" are sequentially fed into the network. The synaptic behaviour was examined through optical potentiation and electrical depression characteristics of our device as shown in Figure 6(b). In this experiment, the device is illuminated with 10 light pulses of wavelength 532 nm, power 60 $\mu$W, $\Delta w = 5$ s and $\Delta t = 5$ s at $V_{ds} = 10$ V and $V_{gs} = 0$ V. The light is then switched off and the electrical depression is carried out by applying $V_{gs}$ pulses of amplitude 0.3 V, $\Delta w = 3$ s and $\Delta t = 3$ s for 10 times. In Figure 6(b), the peak ΔEPSC after every pulse is plotted as a function of the pulse number for both the potentiation and the depression processes. Nonlinearity factors $\alpha$ associated with long term potentiation (LTP) and long term depression (LTD) is obtained by fitting the experimental results shown in Figure. 6(b) with the following equations.[73]

$$G = ([G_{max}^\alpha - G_{min}^\alpha] \times \omega + G_{min}^\alpha)^{\frac{1}{\alpha}} \quad \text{if } \alpha \neq 0 \quad (8)$$

$$= G_{min}^\alpha \times \left(\frac{G_{max}^\alpha}{G_{min}^\alpha}\right)^\omega \quad \text{if } \alpha = 0 \quad (9)$$

Where $G$ represents the peak ΔEPSC after every pulse, while $G_{max}$ and $G_{min}$ stand for the overall maximum and minimum (nonzero) ΔEPSC for either potentiation (LTP) or depression (LTD). $\omega$ is an internal variable. The non linearity factors $\alpha$ is found to be 2.29 and 0.24 for LTP and LTD, respectively. Note that for near ideal device, $\alpha$ must be 1.[73] This is the key factor for determining the image recognition accurecy of the neural network. The synaptic weights in our ANN model are updated using a backpropagation algorithm with cross-entropy loss as the cost function under zero-bias conditions, while a rectified linear unit (ReLU) activation function has been employed for efficient information propagation across layers. Learning rate (fixed) and batch size was $1 \times 10^{-3}$ and 12000, respectively. Training was conducted over 60 epochs, achieving a mean image recognition accuracy of ~85%, as plotted in Figure 6(c). These results demonstrate that our optoelectronic synapse effectively mimics biological synaptic functions and is well-suited for neuromorphic computing, arithmetic operations, and energy-efficient image recognition.

### 3. Conclusion

Bilayer $MoS_2$ based FETs can act as optoelectronic synapses with very high pair pulse facilitation (PPF) index. Two very important characteristics of human brain namely the short term to long term memory (STM-to-LTM) transition and learning-forgetting-relearning behaviour can be replicated in these devices. It has also been found that the rate of depression of the memory state can be controlled by the gate bias. Moreover, the energy consumption per synaptic event in these devices is estimated to be as low as 280 fJ (20 nJ) for electrical (optical) stimuli. These artificial synapses can perform arithmetic and image recognition operations. Artificial neural network (ANN) simulation of these devices shows an image recognition accuracy of ~85%. All these results underscore the potential of $MoS_2$ for the power-efficient, transparent and flexible artificial synaptic devices for the future application of neuromorphic computing.

### Acknowledgement

Author acknowledges the financial support from the Science and Engineering Research Board (SERB), under the grant number CRG/2022/001852, Government of India. We thankful to Industrial Research and consultancy Centre (IRCC), Sophisticated Analytical Instrument Facility (SAIF), and the Centre for Excellence in Nanoelectronics (CEN), IIT Bombay for providing the characterization facilities.

### Conflict of interest

The author declares no conflict of interest.

### Data availability statement

Data that support the findings of this study are available from the corresponding author upon reasonable request.

Supporting information for

# CVD grown bilayer MoS2 based artificial optoelectronic synapses for arithmetic computing and image recognition applications

Umakanta Patra[1], Subhrajit Sikdar[1,2], Roshan Padhan[3], Amandeep Kaur[1], Satyaprakash Sahoo[4,5], Subhabrata Dhar[1*]

[1]Department of Physics, Indian Institute of Technology, Bombay, Powai, India.

[2]International Center for Materials Nanoarchitectonics (MANA), National Institute for Materials Science, Tsukuba, Ibaraki 305-0044, Japan.

[3]Layered Materials and Device Physics Laboratory, Department of Chemistry, Physics and Atmospheric Science, Jackson State University, Jackson, MS 39217, USA.

[4] Laboratory for Low Dimensional Materials, Institute of Physics, Bhubaneswar, Odisha 751005, India.

[5] Homi Bhabha National Institute, Training School Complex, Anushakti Nagar, Mumbai, 400094, India.

*Corresponding author email: dhar@phy.iitb.ac.in

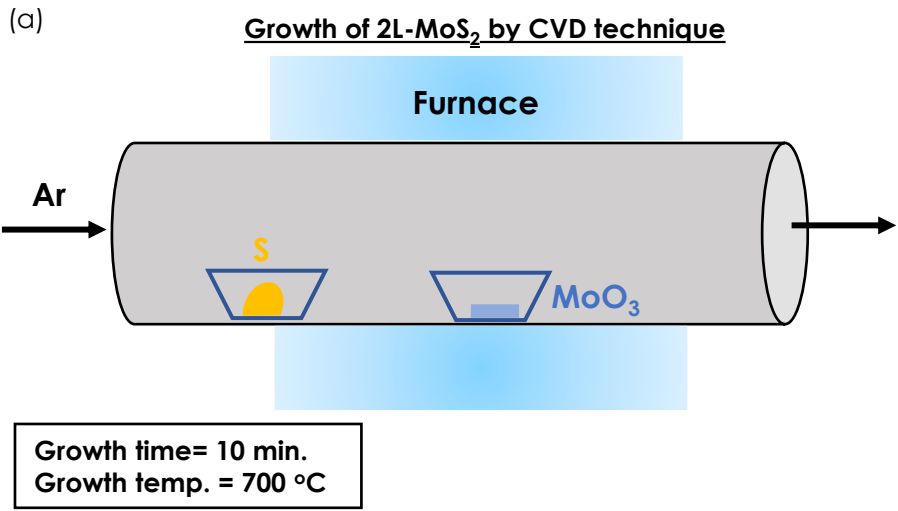

**Figure S1.** (a) Schematic representation of the CVD set-up for the growth of 2L-MoS$_2$.

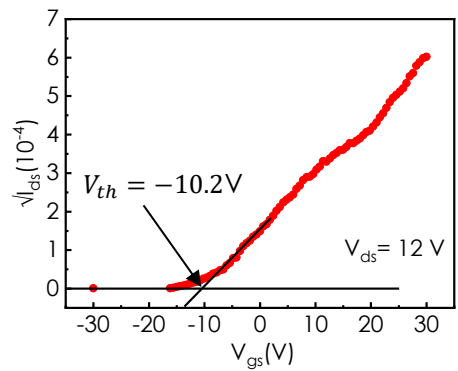

**Figure S2.** (a) Variation of $\sqrt{I_{ds}}$ vs $V_{gs}$ at $V_{ds} = 12$ V.

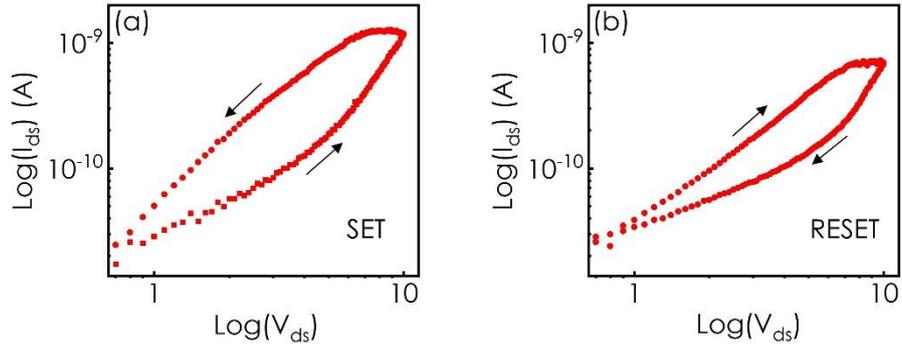

**Figure S3.** Log($I_{ds}$) versus Log($V_{ds}$) plots recorded at $V_{gs} = 0$ V during (a) SET and (b) RESET processes of the device.

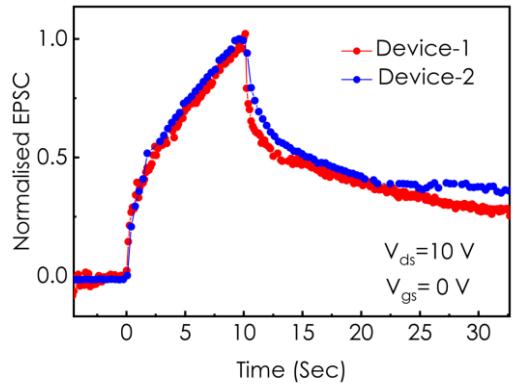

**Figure S4.** Normalised EPSC as a function of time as the light is switched on and off for two devices fabricated on different parts of the same MoS$_2$ bilayer film.

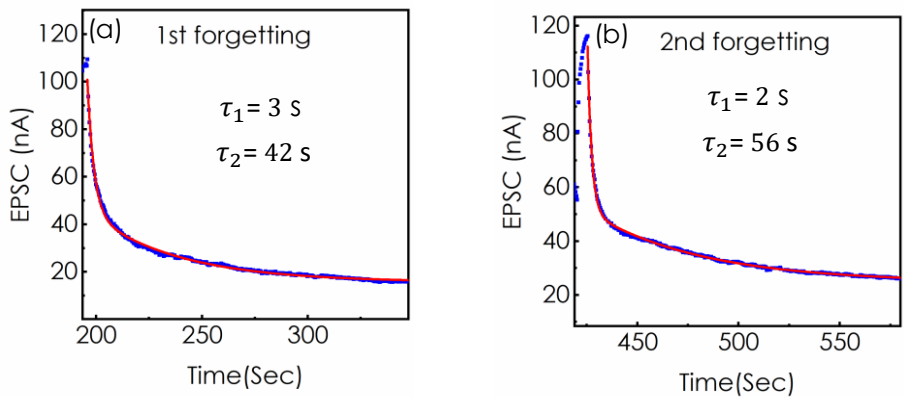

**Figure S5.** Decay profiles for (a) 1$^{st}$ and (b) 2$^{nd}$ forgetting process.

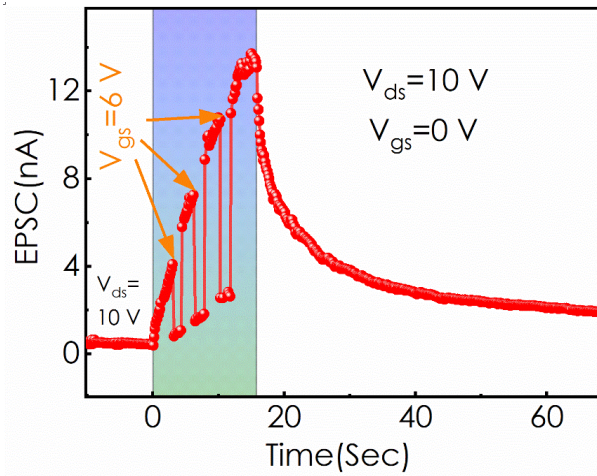

**Figure S6.** Time variation of the EPSC signal. Shaded area represents the illumination (532 nm laser) on time. Arrows mark the time points for the application of the positive gate bias ($V_{gs}$). EPSC has found to fall sharply to a much lower value whenever $V_{gs} = +6$ V is introduced (even under illumination).

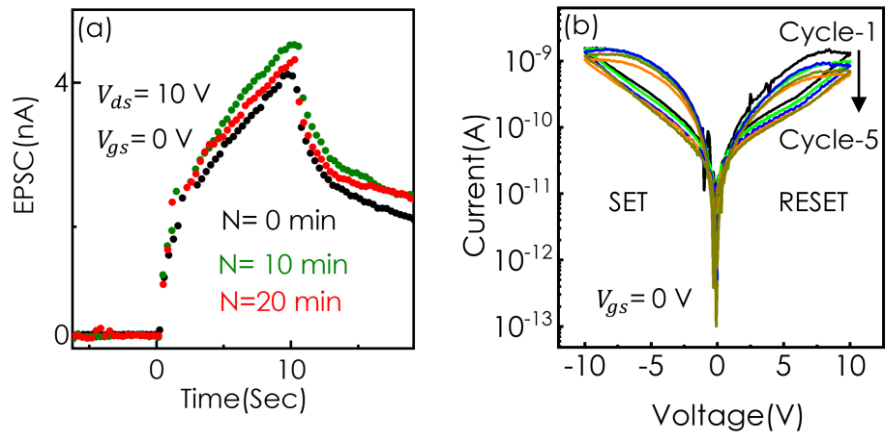

**Figure S7.** (a) Variation of EPSC with time upon switching on and off the light exposure recorded after different time intervals. (b) Current-voltage profiles of the device recorded for five cycles. Figure (a) shows the time variation of EPSC upon switching on and off the light exposure recorded after different time intervals. (b) represents the current-voltage profiles of the device recorded for several cycles. The cycle-to-cycle variation can be estimated as ~ 3%, ~ 11 % (SET) and ~ 20 % (RESET) in EPSC [Figure (a)] and current [Figure (b)] values, respectively. This suggests a good stability of our devices in response to optical and electrical stimuli.

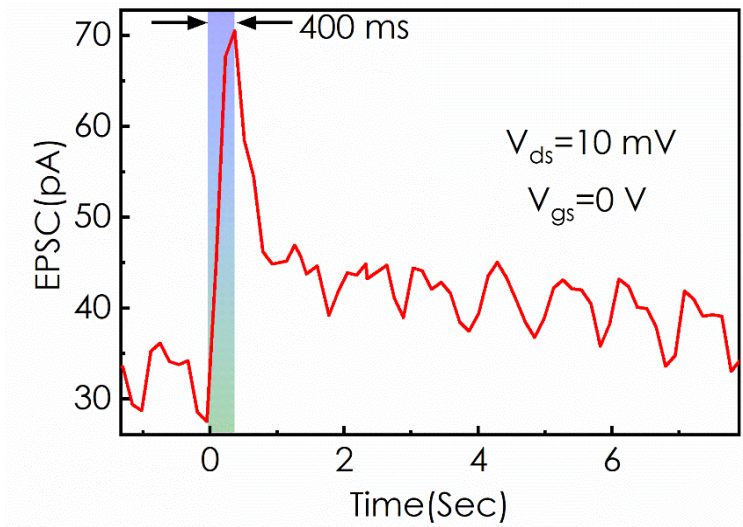

**Figure S8.** Time variation of the EPSC signal. Shaded area represents the illumination (532 nm laser) on time. Electrical energy consumption per synaptic event is estimated to be 280 fJ.